\documentclass[12pt]{article}
\usepackage{amsfonts,amssymb,amsmath,amscd}
\usepackage{amsmath,amssymb,epsfig,ulem}
\newlength{\wth}
\newcommand{\startappendix}{
\setcounter{section}{0}
\renewcommand{\thesection}{\Alph{section}}}

\newcommand{\Tr}{{\rm Tr}}

\newcommand{\RR}{{\mathbb R}}
\newcommand{\CC}{{\mathbb C}}

\newcommand{\ad}{{\rm ad}}

\renewcommand{\Re}{{\rm Re}}

\newcommand{\bz}{{\bar z}}
\newcommand{\bw}{{\bar w}}

\newcommand{\cA}{{\mathcal A}}

\newcommand{\tq}{{\tilde q}}
\newcommand{\cV}{{\mathcal V}}

\newcommand{\cM}{{\mathcal M}}

\newcommand{\cN}{{\mathcal N}}

\newcommand{\cW}{{\mathcal W}}

\newcommand{\dbar}{{\overline{ \partial}}}

\newcommand{\tx}{{\tilde x}}

\newcommand{\vphi}{{\varphi}}
\newcommand{\ot}{\otimes}
\newcommand{\bK}{{\bar K}}
\newcommand{\blambda}{{\bar\lambda}}

\def\be{\begin{equation}}
\def\ee{\end{equation}}
\def\bear{\begin{eqnarray}}
\def\eear{\end{eqnarray}}

\newcommand{\dL}{{\widehat\Lambda}}
\def\half{{ \frac{1}{2} }}

\def\dg{{\dagger}}

\def\Re{{\rm Re\hskip0.1em}}

\def\Id{{\mathbb{I}}}

\def\a{{\alpha}}

\def\dg{{\dagger}}

\def\tC{{{\tilde C}}}

\def\tq{{{\tilde q}}}

\def\bpsi{{\overline{\psi}}}
\def\bchi{{\overline{\chi}}}

\def\delbar{{\overline {\partial}}}

\def\vert{{|}}

\def\tx{{\tilde x}}
\def\ty{{\tilde y}}

\newcommand{\LG}{{{}^LG}}

\begin{document}

\thispagestyle{empty}
\begin{flushright}\footnotesize
\texttt{pi-strings-241}\\
\vspace{2.1cm}
\end{flushright}

\renewcommand{\thefootnote}{\fnsymbol{footnote}}
\setcounter{footnote}{0}

\begin{center}{\bf \Large A note on Wilson-'t Hooft operators}

\vspace{2.1cm}

\textrm{Natalia Saulina}

\vspace{1cm}

\textit{Perimeter Institute, Waterloo, ON 
N2L2Y5, Canada}\\
\texttt{nsaulina, perimeterinstitute.ca}

\bigskip


\par\vspace{1cm}

\textbf{Abstract}\vspace{5mm}
\end{center}

\noindent
We find  the basic ingredients required to compute 
the Operator Product Expansion of Wilson-'t Hooft operators
in $\cN=4$ super-Yang-Mills theory with gauge group $G=PSU(3)$. These include the geometry of certain moduli spaces of BPS configurations in the presence of 't Hooft operators and vector bundles over them.
The bundles arise in computing the OPE due to electric degrees of freedom in dyonic operators. We verify our results by reproducing the OPE of
't Hooft operators predicted by S-duality.

\vspace*{\fill}

\setcounter{page}{1}
\renewcommand{\thefootnote}{\arabic{footnote}}
\setcounter{footnote}{0}

 \newpage

\tableofcontents

\section{Introduction}
The famous S-duality
conjecture \cite{MO} states that $\cN=4$ super-Yang-Mills (SYM) theory
with gauge group $G$ is isomorphic to $\cN=4$ SYM
 with the Langlands-dual  gauge group $\LG.$ This isomorphism maps Wilson loop operators \cite{Wilson}
to 't Hooft loop operators \cite{GNO}, \cite{KWH} in the dual theory. Recalling that the product of Wilson loops is determined by the
representation ring of $G,$ S-duality conjecture predicts that product of 't Hooft loops is controlled by the
representation ring of $\LG.$ This  prediction has been verified
in \cite{KW} based on the earlier mathematical result \cite{Lusztig}.

Yang-Mills theory also admits mixed Wilson-'t Hooft (WH) loop operators.
As explained in \cite{KWH}, at zero $\theta-$angle  they are labeled by elements of the set
$$
\dL(G)/\cW=\Bigl(\Lambda_w(G)\oplus \Lambda_w(\LG)\Bigr)/\cW,
$$
where $\Lambda_w(G)$ is the weight lattice of $G$ and $\cW$ is the
Weyl group (which is the same for $G$ and $\LG$). In $\cN=4$ SYM
theory these mixed operators can be made supersymmetric preserving
 one quarter of the original supersymmetry.

In \cite{KS} we outlined an approach how to compute the product
of WH loop operators for general group $G$ and
actually computed it for $G=SU(2)$ and $G=PSU(2).$ Our approach uses
 the holomorphic-topological twist  \cite{htft} of
the $\cN=4$ SYM theory and the connection between BPS configurations in $\cN=4$ SYM theory 
in the presence of 't Hooft operators and solutions of 3d Bogomolny equations 
with magnetic sources \cite{KW},\cite{Witten2009}.
For $G=SU(2)$ and $G=PSU(2)$
S-duality completely fixes the OPE of Wilson-'t Hooft loop operators
and our results were in agreement with S-duality.

More recently alternative methods of computing OPEs of loop operators
in a certain class \cite{Gaiotto} of $\cN=2$ SYM theories were proposed
\cite{AGGTV},\cite{DGOT},\cite{DGOT2},\cite{GMN},\cite{T} using connection with 2d Conformal Field Theory \cite{AGT}. The algebra of WH loop operators for gauge groups $SU(2)$ and $PSU(2)$ can be explicitly
determined using these references.

For other gauge groups very little is known about the algebra of loop
operators.  There is some partial information \cite{Gomis} arising from
conjectural connection with Toda CFT.
  As we clarify in Section \ref{rev}, our approach \cite{KS}
works for general $\cN=2$ theories. This gives an opportunity to
compute OPEs of loop operators in these theories in our approach
and compare with the forthcoming results from the alternative methods \cite{GMN2}. For example, it is interesting to  find the complete algebra of loop operators in $\cN=2$ theories with gauge group $PSU(n)$
for $n>2.$

The simplest non-trivial OPE of WH operators in $\cN=4$ SYM theory for  $G=PSU(3),$  which is not predicted by  S-duality, is
\be \label{goal}WT_{\mu,\nu}\times
WT_{\mu,0}= WT_{2\mu, \, \nu}+\sum_j (-)^{s_j}\, WT_{\overline{\mu}, \,\nu_j}\ee
where magnetic charge $\mu=w_1$($\overline \mu=w_2$) is the highest weight of
a fundamental (anti-fundamental) representation
of $\LG=SU(3)$ and electric charge
$\nu=aw_1+bw_2$ is the highest weight of $G,$ i.e. $a+2b=0$ mod $3.$
The  electric weights $\nu_j$ and signs $(-)^{s_j}$ on the right side of (\ref{goal}) are to be determined.
As we already noted in \cite{KS}, there could be minus signs on
the right side of OPEs arising for the following reason.
Loop operators can be
promoted to line operators.
While loop operators form a commutative ring, line operators form a
monoidal category. We argued in \cite{KS} that the ring of loop operators can be
thought of as the $K^0$-group of this category 
and in K-theory negative signs occur naturally.

As we review in Section \ref{rev}, to compute (\ref{goal}) in our approach,
 we first need to determine  the geometry of the moduli space $\mathcal{M}$ of 3d Bogomolny equations in $I\times \mathcal{C}$ with two sources, each
characterized by magnetic charge $\mu.$ Here $I$ is an interval and $\mathcal{C}$ is a Riemann surface, and boundary conditions at the two
ends of $I$ are such that without any magnetic sources there is unique vacuum.
$\mathcal{M}$ is obtained by blowing-up
certain singular 4-fold $X_4$ which is the compactification of the moduli space of solutions of 3d Bogomolny equations in $I\times \mathcal{C}$ with a single source
characterized by magnetic charge $2\mu.$ The blow-up procedure produces exceptional divisor $D$ in $\mathcal{M}.$
We further must write the appropriate  metric on
the bulk part  $\mathcal{M}_{bulk}$ which is obtained by removing
from $\mathcal{M}$ the vicinity of $D$, i.e. the total space of the normal bundle of $D$ in  $\mathcal{M}.$

The next step to determine the right side of (\ref{goal}) is to find vector
bundles $\mathcal{V}$ over $\mathcal{M}$ and
$\mathcal{V}_{bulk}$ over $\mathcal{M}_{bulk}.$
These bundles arise in computing the OPE due to electric degrees of freedom in dyonic operators.
Equipped with these vector bundles, one should compute cohomology groups
$$H^p(\mathcal{M},\Omega^q\otimes \mathcal{V})\quad \quad \text{and} \quad \quad H^p(\mathcal{M}_{bulk},\Omega^q\otimes \mathcal{V}_{bulk}).$$
 For compact space $\mathcal{M}$
these are sheaf cohomology groups but for non-compact $\mathcal{M}_{bulk}$
we are interested in $L^2$ Dolbeault cohomology of the corresponding
bundles.
From these cohomology groups one will be able
to determine the right side of equation (\ref{goal}).
Namely the first term in (\ref{goal}) comes from the bulk part of the
moduli space
while the second sum is the so called bubbled contribution accounted by
 \be \label{bub_intro}\sum_{p,q}(-)^{p+q} \Biggl(H^p(\mathcal{M},\Omega^q\otimes \mathcal{V})-H^p(\mathcal{M}_{bulk},\Omega^q\otimes \mathcal{V}_{bulk})\Biggr).
 \ee
 The existence of bubbled contribution is due to monopole
 bubbling \cite{KW} which occurs when the magnetic charge of the 't Hooft
 operator  decreases by absorbing a BPS monopole. This process is
 possible because the moduli space of solutions of 3d Bogomolny equations in the presence of
 magnetic source with charge $2\mu$ is non-compact.
For gauge group $G=PSU(2)$ it was possible to
write a complete $PSU(2)$ invariant metric on the ``bubbled geometry"
 and compute cohomology corresponding to
the bubbled  contribution using this metric \cite{KW},\cite{KS}.
We found that for $G=PSU(3)$ there is no complete
$PSU(3)$ invariant metric on the ``bubbled geometry".
For this reason we adopt the procedure outlined above.

This note is organized as follows. In Section \ref{rev} we review our approach.
 We determine the geometry of  $\mathcal{M}$  in Section \ref{total}
and of
$\mathcal{M}_{bulk}$   in Section \ref{bulk}.  As a  non-trivial check of our results, in Section \ref{check} we prove explicitly the OPE
of 't Hooft operators which is expected from S-duality:
\be \label{answer}WT_{\mu,0}\times WT_{\mu,0}=WT_{2\mu,0}+WT_{\bar \mu,0}.\ee
We compute $L^2$ Dolbeault cohomology of $\cM$ in Appendix A and of $\cM_{bulk}$
in Appendix B and in Section \ref{check} we show explicitly how principle $SU(2)$ subgroup of $\LG=SU(3)$
acts on the cohomology of $\cM$ and $\cM_{bulk}$ in agreement with general facts about  moduli spaces
of BPS configurations in the presence of 't Hooft operators
\cite{Witten2009}.
Finally, in Section \ref{bundle}, we construct bundles $\mathcal{V}$
over $\mathcal{M}$ and $\mathcal{V}_{bulk}$
over $\mathcal{M}_{bulk}$ corresponding to $\nu=aw_1+b w_2$ with
$a+2b=0 \, \text{mod}\, 3.$
We will use the geometry of $\cM$ and $\cM_{bulk}$ together with these bundles   to determine the right side of
(\ref{goal}) in $\cN=4$ SYM theory and in $\cN=2$ SYM with
$N_f=0$ in the future \cite{S}.

\section{OPE of Wilson-'tHooft operators in $\cN=4$ SYM: Review}
\label{rev}
In \cite{KS} we outlined an approach to study the Operator Product Expansion of Wilson-'t Hooft operators
in $\cN=4$ SYM theory with gauge group $G$.
The key step in our approach is to  use holomorphic-topological twist \cite{KWH} of $\cN=4$ SYM theory on a manifold $C\times \Sigma$ where  $C$ and $\Sigma$  are Riemann surfaces.  It is convenient to treat $\cN=4$ SYM as $\cN=2$
SYM with a hypermultiplet in the adjoint representation. The theory
has $SU(2)_R\times U(1)_N\times U(1)_B$ symmetry. The holonomy group
is $U(1)_C\times U(1)_\Sigma$. One twists $U(1)_C$ action by a
suitable linear combination of $U(1)_R\subset SU(2)_R$ and $U(1)_B$,
and twists $U(1)_\Sigma$ by $U(1)_N$. The twisted theory is holomorphic-topological
in a sense that correlators of various operators depend holomorphically on
insertion points on $C$ and are completely independent of positions
of the operators
on $\Sigma.$
The field content of the twisted theory depends on complex structures of $C$ and $\Sigma$. However, the dependence on the complex
structure on $\Sigma$ can be eliminated \cite{KWH}.

Let $w$ and $z$ be a complex coordinate on $\Sigma$ and $C$
correspondingly.
The twisted field theory has the following bosonic fields: the
gauge field $A$, the adjoint Higgs field $\vphi=\Phi_w dw\in
K_\Sigma\ot \ad(E)$, the adjoint Higgs field $q=q_\bz d\bz\in
\bK_C\ot \ad(E),$ and the adjoint Higgs field $\tq\in \ad(E)$. Here
$K_\Sigma$ and $K_C$ are the pull-backs of the canonical line
bundles of $\Sigma$ and $C$ to $\Sigma\times C$. We also define
$\Phi_\bw=\Phi_w^\dag$ and $q_z=q_\bz^\dag$.

The fermionic fields are the ``gauginos''
$\lambda_w,\blambda_\bw,\lambda_z, \blambda_z, \lambda_{\bz w},
\blambda_{\bz \bw}, \lambda_{w\bw},\blambda_{w\bw}$ and the
``quarks'' $\psi_\bw, \bchi_w, \psi_\bz, \bchi_\bz, \chi_{z\bw},
\bpsi_{zw}, \chi_{z\bz}, \bpsi_{z\bz}.$ The fermions are all in the
adjoint representation.

Let us recall how BRST-invariant loop operators look like in the twisted
theory. If $\gamma$ is a
closed curve on $\Sigma$ and $p$ is a point on $C$, the BRST invariant Wilson
operator has the form:
$$
W_R(\gamma,p)=\Tr_R\, P\exp i \int_{\gamma\times p}\cA
$$
Here $\cA_w=A_w+i\Phi_w$ and $\cA_{\bar w}=
A_{\bar w}+i\Phi_{\bar w}.$

Next,  the BRST invariant 't Hooft operator $WT_{\mu,0}$ is a disorder operator
prescribing the following singular behaviour for the fields near
the support $\gamma$ which can be locally written as
$\Re w=0,\, z=0$:
\begin{equation}\label{HopF}
F\sim \star_3 d\left(\frac{\mu}{2r}\right)\quad \Phi_w\sim \frac{\mu}{2r}.
\end{equation}
where $\mu$ is in the Lie algebra of the gauge group and $r^2=\vert z\vert^2+ (\Re w)^2$ locally near $\gamma.$ 

Finally, there are mixed BRST invariant Wilson-'t Hooft loop operators
which source both electric and magnetic fields. To describe them, one
requires the singularity of fields as in (\ref{HopF}) and inserts into the
path-integral a factor
$$
\Tr_{R}\, P\exp i \int_{\gamma\times p}\cA
$$
where $R$ is an irreducible representation of the stabilizer
subgroup\footnote{$G_{\mu}=\{g=e^t \in G: [t,\mu]=0\}.$}  $G_\mu\subset G$ of $\mu$.

Note that holomorphic-topological twist is well-defined for any $\cN=2$ super-conformal
gauge theory for arbitrary choice of $C$ and $\Sigma.$
 To compute the OPE of a pair of Wilson-'t Hooft line operators
we actually take $\Sigma=I\times
\RR$ where $I$ is an interval. In this case $\Sigma$ is flat and one
does not twist along $\Sigma.$
Hence one does not need the existence of non-anomalous $U(1)_N$
symmetry for $\Sigma=I\times
\RR$ and can apply our method of computing OPE to general $\cN=2$
gauge theories.

To compute the OPE of a pair of Wilson-'t Hooft line operators we followed the same method as in \cite{KW}. Namely, we quantize the
twisted gauge theory on a manifold  $C\times I\times
\RR$, with suitable boundary conditions and with two insertions of
Wilson-'t Hooft operators that are
sitting at the same point on $C$.
The problem
reduces to the supersymmetric quantum mechanics on the space of zero
modes of the gauge theory.

The twisted theory is independent of gauge coupling \cite{KWH}
and semiclassical computation (at weak coupling) is exact.
 When
quantizing the theory at weak coupling, the roles of Wilson and 't
Hooft operators are very different. 't Hooft operators directly
affect the equations for the BRST-invariant configurations whose
solutions determine the moduli space, i.e. the space of bosonic zero modes. A Wilson
operator corresponds to inserting an extra degree of freedom, which
couples weakly to the gauge fields, and can be treated
perturbatively.

As in \cite{KW}, we choose boundary conditions so that in
the absence of Wilson-'t Hooft line operators the Hilbert space of
the twisted gauge theory is one-dimensional. For explicit choice
of such boundary conditions, see sections 5.2 and 5.3 in \cite{KS}.
Let $\mathcal{M}$ be the moduli space of BPS configurations in the presence 
 of 't Hooft operators. As shown in \cite{KW} and \cite{Witten2009},  $\mathcal{M}$ is
 the moduli space of solutions of 3d Bogomolny equations with magnetic
 sources and
 can also be identified with the moduli space of Hecke modifications of a holomorphic  vector bundle on $C.$
  The type of modifications is determined in terms of magnetic charges
  of 't Hooft operators.
 We showed in \cite{KS} that in $\cN=4$ SYM theory, after holomorphic-topological twist, ``gaugino'' zero modes  span
anti-holomorphic tangent bundle  $\overline{T}\mathcal{M},$
meanwhile ``quark'' zero modes span holomorphic tangent bundle
$T\mathcal{M}.$ Therefore the Hilbert space of the
effective SQM is the space of $L^2$ sections of the vector bundle
\be \label{hilb0}
\oplus_p \Lambda^p\left(T^*\cM\oplus {\overline
T^*}\cM\right)=\oplus_{p,q}\Omega^{p,q}(\cM).
\ee
We also showed that  the BRST operator acts as the
Dolbeault operator.

When electric degrees of freedom are switched on, i.e. one  inserts
Wilson-'t Hooft operators as opposed to 't Hooft operators, one
has to read off vector bundle $\mathcal{V}$ over $\mathcal{M}$ from the electric
charges of the operators. Then the Hilbert space of the
effective SQM is the space of $L^2$ sections of the vector bundle
\be \label{hilb}
\mathcal{V}\ot \Bigl(\oplus_{p,q}\Omega^{p,q}(\cM)\Bigr).
\ee
and the BRST operator acts as the covariant
Dolbeault operator.

To compute OPE of Wilson-'t Hooft operators $WT_{\mu_1,\nu_1}WT_{\mu_2,\nu_2}$ (for $\mu_1$ and $\mu_2$ - minuscule\footnote{Weights of
a minuscule representation form a single Weyl orbit. Moduli spaces of Hecke modifications corresponding to minuscule magnetic weights are smooth and compact.}
representations of
group $\LG$) in our approach, one should first find the moduli space
$\mathcal{M}_{\mu_1+\mu_2}$ corresponding to magnetic charge $\mu_1+\mu_2$. This space is non-compact and its compactification results
in a singular manifold. Resolving the singularity one gets compact manifold
$\cM$. One should further excise the vicinity of the blown-up
regions to get non-compact manifold $\cM_{bulk}.$ The next step
is to construct vector bundles $\mathcal{V}$ on $\cM$ and $\mathcal{V}_{bulk}$ on $\cM_{bulk}.$ The information about these bundles
is encoded in electric weights $\nu_1,\nu_2.$
Then the bulk
contribution to the zero mode Hilbert space is given by
$\sum_{p,q}(-)^{p+q}H^p(\mathcal{M}_{bulk},\Omega^q\otimes \mathcal{V}_{bulk}),$
while the bubbled contribution is captured by
\be \label{res_bub}\sum_{p,q}(-)^{p+q} \Biggl(H^p(\mathcal{M}_{total},\Omega^q\otimes \mathcal{V})-H^p(\mathcal{M}_{bulk},\Omega^q\otimes \mathcal{V}_{bulk})\Biggr).\ee

To compute the OPE in $\cN=2$ SYM theory with $N_f=0$ we simply
note that ``quark" zero modes are absent so in the Hilbert spaces
(\ref{hilb0}) and (\ref{hilb})
the sum goes only over $p$ with $q=0.$ Therefore,
the bubbled contribution is given by the sum similar to (\ref{res_bub}) but with $q=0$

\section{Geometry  of $\mathcal{M}$}
\label{total}
Let us take gauge group $G=PSU(3).$ In this section we find the moduli space  $\mathcal{M}$
of BPS  configurations in $\cN=4$ SYM theory on $R\times I\times \mathcal{C}$ with two
't Hooft operators $W_{\mu,0}$ inserted at points in $I\times \mathcal{C}.$
Here $\mu=w_1$ is the highest weight of the fundamental representation
of $\LG=SU(3).$
As reviewed in Section \ref{rev}, the boundary conditions at the ends of the interval $I$ are chosen 
such that there is unique vacuum in the absence of 't Hooft operators.

Recall that the insertion of a t-Hooft operator can be viewed as a Hecke modification of a
holomorphic vector bundle on $\mathcal{C}$ \cite{KW},\cite{Witten2009}.
Let us first find out moduli space of Hecke modification corresponding to
'tHooft operator $WT_{2\mu,0}.$
Starting from holomorphic vector bundle $E_-$ with sections
 $$s_a(z)=\bigl(e_a(z),f_a(z),h_a(z)\bigr) \quad a=1,2,3$$
 the Hecke modification corresponding to $2\mu$ produces vector bundle $E_+$ with sections:
 $$s_1'=z^{-2}s_1,\quad s_2'=s_2,\quad s_3'=s_3.$$
Let us expand $e_1(z)=u^1+zv^1,\quad f_1(z)=u^2+zv^2,\quad h_1(z)=u^3+zv^3$
where non-degeneracy requires that $u^1,u^2,u^3$ cannot vanish simultaneously.
In this way we find the general local holomorphic section of $E_+$:
$$s(z)={c(u^1+zv^1,u^2+zv^2,u^3+zv^3)\over z^2}+{d(u^1,u^2,u^3)\over z}+\ldots$$
where $\ldots$ stands for holomorphic functions and $c,d$ are complex numbers.

There are identifications on parameters of the Hecke modification:\be\label{constr_i}(u^1,u^2,u^3,v^1,v^2,v^3)\sim t (u^1,u^2,u^3,v^1,v^2,v^3) \quad t \in \CC^*\ee
\be\label{constr_ii}v^a\sim v^a+wu^a \quad w\in \CC^*\ee
The three invariant combinations under (\ref{constr_ii}) are
$$y_a=\epsilon_{abc}u^bv^c.$$
These have weight 2 under (\ref{constr_i}) and  satisfy
\be \label{hyp} u^a y_a=0.\ee
So we conclude that moduli space of the Hecke modification corresponding to $WT_{2\mu,0}$ is
given by a hypersurface (\ref{hyp}) in the weighted projective space $W_{111222}$
with coordinates $u^1,u^2,u^3,y_1,y_2,y_3$ and with $u^1=u^2=u^3=0$ locus excluded (so that this moduli space is non-compact).

To find the bubbled contribution, we  first compactify
this moduli space by adding the locus $\vec u=0.$ The resulting space, let us denote it by $X_4,$
is singular  since  the ambient $W_{111222}$ is singular at  $\vec u=0$
and hypersurface (\ref{hyp}) passes through
this singularity. In fact $X_4$ near $\vec u=0$ looks like $\mathbb{C}^2/Z_2$ fibered over $\mathbb{P}^2_{\vec y}.$

We resolve this singularity
by blowing up $W_{111222}$
$$u^au^b=\Lambda U^aU^b$$ with homogenous coordinates $U^a$ on the exceptional $\mathbb{P}_{\vec U}^2.$ The weights under the two $\mathbb{C}^*$ actions are
\be \label{action}\begin{array}{c|c|c|c|c|c|c|c}
&U^1 & U^2 & U^3 & \Lambda & y_1 & y_2 & y_3\cr
\hline
\text{new}&1 & 1 & 1& -2& 0& 0& 0 \cr
\hline
\text{old}&0& 0& 0& 1& 1&1&1\cr
\end{array}
\ee

Note that blow-up of $W_{111222}$ gives 5-fold $Y_5$ which is $\mathbb{P}^3$ fibration over
$\mathbb{P}^2.$ Here $U^1,U^2,U^3$ are homogenous coordinates on the base $\mathbb{P}_{\vec U}^2$
and $\Lambda, y_1,y_2,y_3$  are homogenous coordinates on the
fiber $\mathbb{P}^3.$ An exceptional divisor $\mathcal{D}$ in $Y_5$
is given by $\Lambda=0$ and has topology of
$\mathbb{P}^2_{\vec y} \times \mathbb{P}^2_{\vec U}.$

The proper transform  $\mathcal{M}$ of $X_4$
under this blow-up is $\mathbb{P}^2$ fibration over $\mathbb{P}^2_{\vec U}.$ The 4-fold $\cM$ is defined by $y_aU^a=0$ in the 5-fold $Y_5.$
The exceptional divisor $\mathcal{D}$ intersects $\cM$ over 3-fold $D$
which is
$\mathbb{P}^1$ fibration over $\mathbb{P}^2_{\vec U}.$

We can write the most general $PSU(3)$ invariant
K\"ahler form on $\mathcal{M}$ as
\be \label{metric_tot} (-i)J=f_1(s) \mathcal{E}_1\wedge \overline{\mathcal{E}}_1 +f_2(s)\mathcal{E}_2\wedge
\overline{ \mathcal{E}}_2+ f_3(s){\mathcal E}_3\wedge \overline {\mathcal E}_3+
f_4(s) \mathcal{E}_4 \wedge \overline{  \mathcal{E}}_4 \ee
 where $f_i(s)$
are functions of $PSU(3)$ invariant $s$ (which is also invariant
 under $\mathbb{C}^*\times  \mathbb{C}^*$ action):
 $$s={\vert \Lambda \vert^2 Y^2\over X}, \quad X=y_a \bar y^a, \quad Y={\overline U}_a U^a.$$
 In (\ref{metric_tot}) we used
 \be\label{differ}\mathcal{E}_1=\partial s \quad  \mathcal{E}_2={y_adU^a\over \sqrt{XY}}\quad \mathcal{E}_3={\epsilon^{acd}{\overline U}_ay_c dy_d\over X \sqrt{Y}}
 \quad \mathcal{E}_4=
 {\epsilon_{abc}\bar y^a U^b dU^c\over Y\sqrt{X}}\ee
  It is implied that $\mathcal{E}_i$ are evaluated on a hypersurface $y_aU^a=0.$
  Since ${U^a\over \sqrt{Y}}$ and ${y_a\over \sqrt{X}}$ are multiplied by
 phases under the corresponding  $C^*$ actions, we note that
 \be \label{proper}
 \mathcal{E}_1 \in \Gamma\Bigl(\Omega^{1,0}(0,0)\Bigr),\,\,
 \mathcal{E}_2 \in \Gamma\Bigl(\Omega^{1,0}(1,1)\Bigr),\,\,
 \mathcal{E}_3 \in \Gamma\Bigl(\Omega^{1,0}(-1,2)\Bigr),\,\,
 \mathcal{E}_4 \in \Gamma\Bigl(\Omega^{1,0}(2,-1)\Bigr).\ee
 Here $\Omega^{1,0}(b,f)$ stands for $(1,0)$ form with
 charges $b$ and $f$ under $C^*$ acting on the base and the fiber of the ambient 5-fold $Y_5$ respectively. The K\"ahler form must be invariant under
 the two $C^*$ actions. This is why there are no mixed
 terms such as $\mathcal{E}_i \wedge \overline{\mathcal{E}}_j \, \text{with}\,\, i\ne j$ in the general expression  (\ref{metric_tot}).

Recall that the exceptional divisor $D$ in $\mathcal{M}$ is defined by $\Lambda=0$.
  At $\Lambda=0,$ let us note that $y_a$ are homogenous coordinates on
 $\mathbb{P}^2_{\vec y}.$
   The K\"ahler form
  on $D$
 \be\label{volume_D}J_{D}=C_1J^{FS}_{\mathbb{P}^2_{\vec U}}+
 C_2J^{FS}_{\mathbb{P}^2_{\vec y}}\vert_{y_aU^a=0}\ee
 can be expressed in terms of   \footnote{Evaluated at $\Lambda=0$ i.e.
 treating $y_a$ as homogenous coordinates on $\mathbb{P}^2_{\vec y}$} $\mathcal{E}_i$ using

$$ (-i)J^{FS}_{\mathbb{P}^2_{\vec U}}=\mathcal{E}_2 \wedge
\overline{ \mathcal{E}}_2+\mathcal{E}_4 \wedge
\overline{ \mathcal{E}}_4, \quad (-i)J^{FS}_{\mathbb{P}^2_{\vec y}}=\mathcal{E}_2 \wedge
\overline{ \mathcal{E}}_2+ \mathcal{E}_3 \wedge
\overline{ \mathcal{E}}_3.$$
Let us work in the patch $U^1\ne 0,\,\, y_3\ne 0$ and use
inhomogenous coordinates
$$z_1={U^2\over U^1},\quad z_2={U^3\over U^1},\quad v={y_2\over y_3},\quad \lambda={\Lambda (U^1)^2\over y_3}.$$
The hypersurface equation $y_aU^a=0$ is solved in this patch
as
$$y_1=-y_3(z_2+vz_1).$$
We will set $y_3=1, U^1=1$ to simplify the formulas in the rest
of Section \ref{total}.
In this patch we write $s={\vert \lambda\vert^2 y^2\over x}$ with
$$y=1+\vert z_1\vert^2+\vert z_2\vert^2,\quad x=1+\vert v\vert^2+\vert z_2+v z_1\vert^2.$$
We find differentials (\ref{differ}) explicitly in this patch:
\be \label{differ_ii}\mathcal{E}_1=s \Biggl({d\lambda \over \lambda}-{\partial x \over x}+2{\partial y\over y}\Biggr) \quad
\mathcal{E}_2={(vdz^1+dz^2)\over\sqrt{xy}}\ee
 $$\mathcal{E}_3=-{\sqrt{y}\over x}\Bigl(dv+{(\bar z_1 -v \bar z_2)\over y}(vdz^1+dz^2)\Bigr)\quad
\mathcal{E}_4={1\over y\sqrt{x}}\Bigl( \overline{\a}_2 dz_1 -  \overline{\a}_1dz_2\Bigr )$$
where  we denote:
$$\overline{\a}_1=\bar v(1+\vert z^1\vert^2)+z^1\bar z^2, \quad
\overline{\a}_2=1+\vert z^2\vert^2+\bar v \bar z^1 z^2.$$
Covariant holomorphic differential $\nabla$ acts on
$\omega \in \Gamma\Bigl(\Omega^{(1,0)}(p,q)\Bigr)$ as
$$\nabla \omega=\left(\partial-{p\over 2}\frac{\partial y}{y}
-{q\over 2}\frac{\partial x}{x}\right)\omega$$

 We find:
 \be \label{der0}\nabla \mathcal{E}_1=0, \quad
 \nabla \overline{\mathcal{E}}_1={1\over s}\mathcal{E}_1 \wedge
  \overline{ \mathcal{E}}_1+ s\mathcal{E}_2 \wedge \overline{ \mathcal{E}}_2-
  s\mathcal{E}_3 \wedge \overline{ \mathcal{E}}_3+2s\mathcal{E}_4 \wedge \overline{ \mathcal{E}}_4\ee
  \be \label{der1}\nabla \mathcal{E}_4=0,\quad  \nabla \overline{\mathcal{E}}_4=
\mathcal{E}_3\wedge\overline{\mathcal{E}}_2,
  \quad \nabla \mathcal{E}_2= -\mathcal{E}_3\wedge \mathcal{E}_4\ee
  \be \label{der2}\nabla \mathcal{E}_3=0, \quad
 \nabla \overline{\mathcal{E}}_3=-\mathcal{E}_4 \wedge \overline{\mathcal{E}}_2,  \quad \nabla \overline{ \mathcal{E}}_2=0.\ee

From $dJ=0$ we find that $f_2(s),f_3(s),f_4(s)$ are determined
(up to two integration constants)  in terms of $f_1(s)$:
\be \label{metric_fun}f_2=f_3+f_4,\quad  f_3'=-sf_1,\quad f_4'=2sf_1.\ee

We assume the following asymptotics at $s\mapsto 0$:
\be \label{zero}f_1\sim C_0s^{-1}, \quad f_2\sim C_1+C_2+C_0s,\quad f_3(s)\sim C_2-C_0s \quad f_4 \sim C_1+2C_0s\ee
with $C_0>0,\,C_1>0,\, C_2>0$.
 This ensures that at $\Lambda=0$ we find $D$ with
 K\"ahler form (\ref{volume_D}).

Meanwhile, at $s\mapsto \infty$ we must choose
the asymptotics
\be \label{inf}f_1(s) \mapsto \half A'_0 s^{-3},\quad f_2(s) \mapsto C'_0-\half A'_0 s^{-1},\quad
f_3(s)\mapsto \half A'_0 s^{-1},\quad
f_4(s) \mapsto C'_0-A'_0s^{-1}\ee
where $A_0'>0,C_0'>0.$

 This ensures that at $s \mapsto \infty,$ i.e. as we go away from the exceptional divisor, we find flat space
 fibered over $\mathbb{P}^2_{\vec U}$:
$$(-iJ)=C'_0 \left(-iJ^{FS}_{\mathbb{P}^2_{\vec U}} \right) +
{A'_0\over 2 y^2}\sum'_a \left(dw_a-2{\partial y\over y}w_a\right)
\wedge \left(d{\bar w}_a-2{\overline{\partial} y\over y}\bar w_a\right).$$
In this asymptotic regime $\Lambda\ne 0$ so  we may introduce
$w_a={y_a\over \Lambda}$ and $\sum'_a$ means that we use $w_1=-(w_2z^1+w_3z^2).$

For example, we may take the following functions with the right asymptotics:
  \be \label{choice-tot}f_1={C_2\over s(1+s)^2},\quad f_3={C_2\over 1+s},\quad
f_4=C_1+{2C_2 s\over 1+s},\quad f_2=f_3+f_4\ee
We compute $L^2$ Dolbeault cohomology of $\cM$ using (\ref{choice-tot}) in Appendix A. This information is used to verify that we identified
the geometry of $\cM$ correctly (see Section \ref{check}).
The fact that we have two parameters $C_1$ and $C_2$ is justified
since we show in Appendix A.2 that $h^{1,1}(\cM)=2.$


\section{Geometry of $\mathcal{M}_{bulk}$}
\label{bulk}
To describe bulk geometry we work in the patch
$\Lambda \ne 0$ in $\mathbb{P}^3$ fiber and $U^1\ne 0$ in
$\mathbb{P}^2$ base of $Y_5.$ The appropriate inhomogenous
coordinates are
$$t_a={y_a\over \Lambda} \,\, a=1,2,3; \quad z^1={U^2\over U^1},\quad
z^3={U^3\over U^1}.$$
In this patch we solve the equation $y_aU^a=0$ as
$$t_1=-(z^1t_2+z^2t_3).$$
We can write $PSU(3)$ invariant
K\"ahler form on the bulk geometry $\mathcal{M}_{bulk}$ as

\be \label{metric} (-i)J_{bulk}=g_1(s) \mathcal{E}_1\wedge \overline{\mathcal{E}}_1 +g_2(s)\mathcal{E}_2\wedge
\overline{ \mathcal{E}}_2+ g_3(s){\mathcal E}_3\wedge \overline {\mathcal E}_3+
g_4(s) \mathcal{E}_4 \wedge \overline{  \mathcal{E}}_4 \ee
 where $g_i(s)$
are functions of $PSU(3)$ invariant $s$ which in this patch can be written as (we set $U^1=1$)
$$s={y^2\over \tx},\quad \tx=t_a \overline{t}^a,\quad y=1+\vert z_1\vert^2+\vert z_2\vert^2.$$
In this patch (with $U^1=1,\Lambda=1$) the differentials have the form
$$\mathcal{E}_1=s\Bigl({2\partial y\over y}-{\partial \tilde{x}\over \tilde{x}}\Bigr)\quad \mathcal{E}_2={t_2dz^1+t_3dz^2\over \sqrt{\tilde{x}y}}$$
$$\mathcal{E}_3={y^{1/2}\over \tilde{x}}\Bigl(t_2dt_3-t_3dt_2+
{(t_2\bar z_2-t_3\bar z_1)(t_2dz^1+t_3dz^2)\over y}\Bigr)
\quad \mathcal{E}_4={\bar \alpha_2 dz^1-\bar \alpha_1 dz^2\over y \sqrt{\tilde{x}}}$$
where
$$\bar \alpha_1=\bar t^2 +z^1(\bar t^2\bar z_1+\bar t_3 \bar z_2),\quad
\bar \alpha_2=\bar t^3 +z^2(\bar t^2\bar z_1+\bar t_3 \bar z_2).$$

At $s\mapsto \infty$ (away from the blown-up region) the metric on the bulk geometry should coincide
with the metric on total geometry $\mathcal{M}$. This means that
 at $s\mapsto \infty$ we must choose
the asymptotics

\be \label{inf_bulk}g_1(s) \mapsto \half A'_0 s^{-3},\quad g_2(s) \mapsto C'_0-\half A'_0 s^{-1},\quad
g_3(s)\mapsto \half A'_0 s^{-1},\quad
g_4(s) \mapsto C'_0-A'_0s^{-1}\ee
where $A_0'>0,C_0'>0.$

Meanwhile, at $s\mapsto 0$ we choose
\be \label{zero_bulk}
g_1(s)\mapsto \half {\hat A\over s^{3/2}},\quad
g_2(s)\mapsto \hat C+ \hat A s^{1/2},\quad
g_3(s)\mapsto \hat C - \hat A s^{1/2},\quad
g_4(s)\mapsto 2\hat A s^{1/2},\quad
\ee
with $\hat C >0,\quad \hat A >0.$

We note that at $s \mapsto 0$ (corresponding to
$\Lambda \mapsto 0$) we may use
coordinates $y_a$ and $u^a=\sqrt{\lambda} U^a$
( i.e. coordinates  before the blow-up).
Moreover, $y_a$ in this limit are homogenous coordinates on
$\mathbb{P}^2_{\vec y}.$ So that
$$s={\ty^2\over X},\quad X=y_a \bar y^a,\quad \ty=u^a \bar u_a.$$

 We find
 at $s\mapsto 0$
 $$(-i)J_{bulk}={\hat A \over X^{1/2}}\sum'_a (du^a- {\partial X\over 2X}u^a)\wedge
  (d\bar u^a- {\overline{\partial} X\over 2X}\bar u^a)+
\hat C \left(-iJ^{FS}_{\mathbb{P}^2_{\vec y}}\right),$$
 where $\sum'_a$ means that $u^ay_a=0$ is implied.
 This correctly describes $\vec u=0$ region in the moduli space
 before the blow-up, which was given by hypersurface
 $u^ay_a=0$ in $\mathbb{C}^3/\mathbb{Z}_2$ fibered
 over $\mathbb{P}^2_{\vec y}.$

 We can, for example, make the following simple choice of functions
 $g_i(s)$ with the asymptotics (\ref{inf_bulk}) and (\ref{zero_bulk}):
  \be \label{choice-bulk}g_1={\hat A\over 2}{1\over s^{3/2} (1+s)^{3/2}},\quad
 g_2=\hat A \Bigl(1+{s^{1/2}\over \sqrt{1+s}}\Bigr),\quad
 g_3=\hat A \Bigl(1-{s^{1/2}\over \sqrt{1+s}}\Bigr),
 \quad g_4={2\hat A \, s^{1/2}\over \sqrt{1+s}}\ee
so that $\hat C=\hat A$ in (\ref{zero_bulk}) and 
$C'_0=2\hat A, \, A'_0=\hat A$ in (\ref{inf_bulk}).
We compute  $L^2$ Dolbeault cohomology of $\cM_{bulk}$ using (\ref{choice-bulk}) 
in Appendix B. This information is used to verify that we identified
the geometry of $\cM_{bulk}$ correctly (see Section \ref{check}).
The fact that we have only one parameter $\hat A$ is justified
since we show in Appendix B.2 that $h^{1,1}(\cM_{bulk})=1.$

\section{Consistency check}
\label{check}
Let us make consistency check of our results with known
general facts about moduli spaces of BPS configurations in the
presence of 't Hooft operators \cite{Witten2009}.
In Appendix A, we found the following non-zero cohomology groups for the total
moduli space:
$$H^{(0,0)}(\mathcal{M})\simeq H^{(4,4)}(\mathcal{M}) =\mathbb{V}_1,\quad H^{(1,1)}(\mathcal{M})\simeq H^{(3,3)}(\mathcal{M})=\mathbb{V}_1\oplus \mathbb{V}_1,$$
$$H^{(2,2)}(\mathcal{M})=\mathbb{V}_1\oplus \mathbb{V}_1\oplus \mathbb{V}_1.$$
Here $\mathbb{V}_1$ is one-dimensional (singlet) representation of $PSU(3).$
Let us decompose all 9 harmonic forms which serve as basis vectors into
three groups
\begin{itemize}
\item 
$1,\,J_{tot},\,J_{tot}^2,\,J_{tot}^3,\,J_{tot}^4$
\item
$ \omega^{(2)}_{(1,1)}, \, J_{tot}\wedge  \omega^{(2)}_{(1,1)},\, 
J^2_{tot}\wedge  \omega^{(2)}_{(1,1)}$
\item
$\omega^{(p.s.d)}_{(2,2)}$ 
\end{itemize}
where harmonic (1,1) form $\omega^{(2)}_{(1,1)}$ is
orthogonal to the K\"ahler  form 
$$J_{tot}\wedge * \omega^{(2)}_{(1,1)}=0$$
 and
harmonic (2,2) form  $\omega^{(p.s.d)}_{(2,2)}$ is primitive and self-dual
$$J_{tot}\wedge \omega^{(p.s.d)}_{(2,2)}=0\quad \quad \omega^{(p.s.d)}_{(2,2)}=*\omega^{(p.s.d)}_{(2,2)}.$$
This decomposition is consistent with the general fact that cohomology 
should transform in representations of the principle $SU(2)_{principle}$ subgroup
of the dual group $\LG=SU(3)$ \cite{Witten2009}. The K\"ahler form
$J_{tot}$ plays the role of the raising operator
of $SU(2)_{principle}.$ 

In Appendix B, for the bulk geometry we found:
 $$H^{(0,0)}(\mathcal{M}_{bulk})\simeq H^{(4,4)}(\mathcal{M}_{bulk}) =\mathbb{V}_1,\quad H^{(1,1)}(\mathcal{M}_{bulk})\simeq H^{(3,3)}(\mathcal{M}_{bulk})=\mathbb{V}_1,$$
$$H^{(2,2)}(\mathcal{M}_{bulk})=\mathbb{V}_1\oplus \mathbb{V}_1.$$
We decompose all 6 harmonic forms which serve as basis vectors into
two groups
\begin{itemize}
\item 
$1,\,J_{bulk},\,J_{bulk}^2,\,J_{bulk}^3,\,J_{bulk}^4$
\item
$\omega^{(p.s.d)}_{(2,2)} $
\end{itemize}

where harmonic (2,2) form  $\omega^{(p.s.d)}_{(2,2)}$ is primitive and self-dual
$$J_{bulk}\wedge \omega^{(p.s.d)}_{(2,2)}=0\quad \quad \omega^{(p.s.d)}_{(2,2)}=*\omega^{(p.s.d)}_{(2,2)}.$$
This decomposition is again consistent with \cite{Witten2009} and $J_{bulk}$
plays the role of the raising generator of $SU(2)_{principle}.$ Moreover, this corresponds precisely to
the decomposition of representation  $2\mu$ of $\LG=SU(3),$ which appears  in the  't Hooft operator $WT_{2\mu, 0}$ in the right side of the OPE (\ref{answer}),
into representations with spin $j=2$ and $j=0$
 under the principle $SU(2)_{principle} \subset SU(3).$ 

Comparing total and bulk cohomologies, we see that harmonic 
forms, which serve as a basis for
bubbled contribution, are 
$$ \omega^{(2)}_{(1,1)}, \, J_{tot}\wedge  \omega^{(2)}_{(1,1)},\, 
J^2_{tot}\wedge  \omega^{(2)}_{(1,1)}.$$
This is a representation with spin $j=1$ under $SU(2)_{principle}$
which is consistent with the decomposition of representation $\overline{\mu}$ 
of $\LG=SU(3),$ which appears  in the  't Hooft operator $WT_{\overline{\mu}, 0}$ in the right side of the OPE (\ref{answer}), under $SU(2)_{principle}\subset SU(3).$
We conclude that both the cohomology groups in the bulk and the bubbled contribution are in agreement with  S-duality prediction (\ref{answer}).

\section{Vector bundles over $\mathcal{M}$ and $\cM_{bulk}$}
\label{bundle}
In this section we construct the bundles $\mathcal{V}$ over $\cM$ and
$\mathcal{V}_{bulk}$ over $\cM_{bulk}$ which appear in (\ref{bub_intro})
and are required to compute the OPE (\ref{goal}).

Let us first identify the vector bundle $\mathbf{V}_{a,b}$ over the base $\mathbb{P}_{\vec U}^2$ which corresponds to the electric weight
$\nu=aw_1+bw_2$ (with $a+2b=0 \, \text{mod} \, 3$) in 
the Wilson-'t Hooft operator $WT_{\mu,\nu}.$
Recall that $\mu=w_1$ breaks Lie algebra $su(3)$ to $su(2)\oplus u(1)$
and  $\nu$ tells us to look  for a bundle in  representation $R_b$ with the highest weight $b$ (number of boxes in the Young diagram) of $SU(2)$ and with charge $2a+b$ under $U(1)$.

Let us clarify this. We use the Chevalley basis of $su(3)$:
$$[h^1, E^{\pm \alpha_1}]=\pm 2  E^{\pm \alpha_1},\quad
[h^2, E^{\pm \alpha_2}]=\pm 2  E^{\pm \alpha_2},$$
$$[h^1, E^{\pm \alpha_2}]=\mp  E^{\pm \alpha_2},\quad
[h^2, E^{\pm \alpha_1}]=\mp E^{\pm \alpha_1}$$
where states in a given irreducible representation $\nu=aw_1+bw_2$ 
are labelled by eigenvalues of $h^1$ and $h^2$:
$$h^1 \vert a,b\rangle= a\vert a,b\rangle \quad
h^2 \vert a,b\rangle= b\vert a,b\rangle. \quad
$$
Acting on $\mathbf{3},$ we can represent raising operators corresponding to simple roots and Cartan generators of $su(3)$  as
\be \label{rep}E^{\alpha_1}=\begin{pmatrix}0 & 0 &0 \\ 0& 0&0 \\1& 0& 0\\ \end{pmatrix},\,\, h^1=\begin{pmatrix}-1 & 0 &0 \\ 0& 0&0 \\0& 0& 1\\ \end{pmatrix},\,\,
E^{\alpha_2}=\begin{pmatrix}0 & 0 &0 \\ 0& 0&1 \\0& 0& 0\\ \end{pmatrix}\,\,
h^2=\begin{pmatrix}0&0&0\\0& 1 &0\\0&0& -1\\ \end{pmatrix}
\ee
The operators $h^2, E^{\pm \alpha_2}$ generate $su(2)$
part of the Lie algebra of the unbroken group.
Meanwhile, the generator for $u(1)$ is
$$J=(2h^1+h^2),\quad J \vert a,b\rangle= (2a+b)\vert a,b\rangle
$$
Note that the value of $J$ on the weights of $PSU(3)$
is always in $3\mathbb{Z}$ since $a+2b=3n,\, n \in \mathbb{Z}$ implies $2a+b=3m, \, m \in \mathbb{Z}.$

To write a connection on $\mathbf{V}_{a,b},$ we first
find a connection on the principle $SU(2)\times U(1)$ bundle over $\mathbb{P}^2$ from the metric on $SU(3)$ group manifold viewed
as  $SU(2)\times U(1)$ bundle over $\mathbb{P}^2$:

$$ds^2_{SU(3)}=-\half {\rm Tr} \Bigl (g^{-1}dg\Bigr)^2=-\half \Biggl(
{\rm Tr} \Bigl (G^{-1}dG\Bigr)^2+{\rm Tr} \Bigl (dh\,h^{-1}\Bigr)^2+
2{\rm Tr} \Bigl (G^{-1}dG\, dh\, h^{-1}\Bigr)
\Biggr)
$$
with
$$g=Gh \quad h=\mathbf{k} \, \,exp\bigl[i {t J\over 2}\bigr]\quad \mathbf{k}\in SU(2).$$
Here $t\in [0,2\pi]$ is a coordinate on $U(1)$ part of the fiber while
for $\mathbf{k}\in SU(2)$ we define the  forms $(d\mathbf{k})\mathbf{k}^{-1}={i\over 2}\vec \rho \cdot \vec\sigma, $ with $\vec \sigma$ - Pauli matrices.
Let us parametrize $\mathbf{k}$ in terms of the Euler angles:
$$\mathbf{k}=exp\bigl[i {\psi\over 2}\sigma^3\bigr]\, exp\bigl[i {\theta \over 2}\sigma^1\bigr]\,exp\bigl[i {\phi\over 2}\sigma^3\bigr],$$
where $\theta \in [0,\pi],\, \phi \in [0,2\pi], \, \psi \in [0,2\pi].$
Then, the forms are given by
$$\rho^3=d\psi+cos \theta \, d\phi,\quad \rho^1+i\rho^2=e^{-i\psi}(d\theta+
i \,sin \theta \,d\phi).$$


Similar to \cite{Byrd}, we parametrize the coset representative $G$ as
$$G=\begin{pmatrix} 1 & 0 &0 \cr 0 & \bar b & -\bar c \cr 0 & c & b
\end{pmatrix}\begin{pmatrix} cos\Upsilon & sin \Upsilon & 0\cr
 -sin \Upsilon & cos \Upsilon & 0\cr 0 & 0 & 1\end{pmatrix}$$
where $\vert b\vert^2+\vert c\vert^2=1$ and $\Upsilon\in[0,{\pi\over 2}].$
We  use the following parametrization
$$c=sin \Theta e^{i\Phi_1},\quad \bar b=cos \Theta e^{i\Phi_2}$$
where
$\Theta \in [0,\pi/2],\quad \Phi_1 \in [0,2\pi,\quad \Phi_2 \in [0,2\pi]$
and define forms $\vec \xi$ as
$$D^{-1}dD=i \vec \xi \cdot \vec \sigma,\quad
D=\begin{pmatrix} \bar b & -\bar c\cr
c & b\end{pmatrix}.$$

Then we find
$$\xi^3=sin^2 \Theta d\Phi_1+cos^2 \Theta d\Phi_2,\quad
\xi^+:=\xi^1+i\xi^2=-i e^{i(\Phi_1+\Phi_2)}\, \Bigl(d\Theta +i sin \Theta \, cos \Theta
(d\Phi_1-d\Phi_2)\Bigr).$$
The metric on $SU(3)$ in these coordinates is
$$ds^2_{SU(3)}={3 \over 4}(dt+A^0)^2+{1\over 4}\bigl(\rho^3+A^3\bigr)^2+{1\over 4}\vert \rho^{+}+A^{+}\vert^2+ds^2_{base}$$
Here the connection is
$$A^0=-\xi_3 sin^2 \Upsilon,\quad A^3= \xi^3(1+cos^2\Upsilon),
\quad A^+=2\xi^+ cos \Upsilon$$
and the metric on the base
$$ds^2_{base}=(d\Upsilon)^2+sin^2 \Upsilon \,\vert \xi_+\vert^2+
sin^2 \Upsilon cos^2 \Upsilon \xi_3^2.$$

The coordinates $z_1,z_2$ used in Section \ref{total}  are expressed as
$$z_1= c\, tan \Upsilon,\quad z_2=\bar b \,tan \Upsilon.$$
Note that
$$ds^2_{base}=
{(dz_1 \otimes d\bar z_1+ dz_2 \otimes d\bar z_2)\over y}-{\partial y\otimes \delbar y\over y^2}$$
with $y=1+\vert z_1\vert^2+\vert z_2\vert^2.$
The K\"ahler form on the base $J_{base}=\half J_{FS}$ where
$\left[ {J_{FS}\over 2\pi}\right]=H$ with $H$ - the hyperplane class
on $\mathbb{P}^2.$

In terms of $z_1,z_2$ we write the connection on the principle $U(1)\times SU(2)$ bundle over
$\mathbb{P}^2$ as
\be \label{answerii}A^+={2i\bigl(\bar z_1 d\bar z_2-\bar z_2 d\bar z_1\bigr)\over
\sqrt{y}(y-1)},\quad A^3=-{(y+1)\over y(y-1)}{\rm Im} (\partial y),\quad
A^{0}={ {\rm Im}(\partial y) \over y}\ee
Therefore, the  connection
on the  vector bundle $\mathbf{V}_{a,b}$ over
$\mathbb{P}^2_{\vec U}$ corresponding to the electric weight
$\nu=aw_1+bw_2$ is
\be \label{conab} \mathbb{A}=\mathbb{A}_{(1,0)}+\mathbb{A}_{(0,1)},\quad \mathbb{A}_{(0,1)}={i(2a+b)\over 2}{\delbar y\over y}\Id+
A^i T^{R_b}_i, \quad \mathbb{A}_{(1,0)}=\mathbb{A}_{(0,1)}^{\dg}\ee
with $A^iT^{R_b}_i$
- the $SU(2)$ connection in the representation $R_b.$
This bundle  is a tensor product
 $$\mathbf{V}_{a,b}=O(-(2a+b)H)\otimes \tilde{V}_{b}$$
 where $H$ is a hyperplane class in $\mathbb{P}_{\vec U}^2$
and
 the Chern classes of the vector bundle $\tilde V_{b}$ are
\be \label{chernab}rk(\tilde V_b)=b+1, \quad c_1(\tilde
V_{b})=0, \quad c_2(\tilde V_{b})=\Bigl(-{\kappa(R_b)\over 2}
\int_{\mathbb{P}^2}{F^i\wedge F_i \over (2\pi)^2}\Bigr)H^2.\ee 
Here we defined $\kappa(R)$ as
$$Tr_{R}T^iT^j=\kappa(R) \delta^{ij},\quad \text{so that}\quad  \kappa(R_1)=\half.$$
We compute
$$F^3=dA^3+{i\over 2}A^-\wedge A^+,\quad F^+=dA^++iA^-\wedge A^3$$
and
$$\int_{\mathbb{P}^2} F^i\wedge F_i=-24 \int sin^2 \Upsilon \, cos \Upsilon  d\Upsilon
\int \xi^1\wedge \xi^2\wedge \xi^3=-4(2\pi)^2.$$ Hence,
$$c_2(\tilde V_{b})=2\kappa(R_b)H^2.$$

Since the bundle $\tilde{V}_b$ is the symmetric tensor product $S^b \tilde{V}_1,$ it is crucial to understand the holomorphic structure of $\tilde{V}_1.$  
Let us write the (0,1) part of the connection on $\tilde V_{1}$ as
$$\mathbb{A}^{\tilde V_{1}}_{(0,1)}=i\left( \delbar \mathcal{G}\right)\mathcal{G}^{-1}\quad \quad \mathcal{G}=\begin{pmatrix}\alpha & \alpha\cr
\alpha^{-1}\beta & \alpha^{-1}(1+\beta)\end{pmatrix}$$
where
$$\alpha={y^{1/4}\over (y-1)^{1/2}},\quad \beta=-{{\bar z}_1\over (y-1)z_2}.$$
Let us use $\mathcal{G},$ the transformation matrix from holomorphic to unitary gauge, to compute the norm of a section
$$\psi_{unit}=\mathcal{G}\psi_{hol},\quad
\psi_{hol}=\begin{pmatrix}\psi_1\cr \psi_2
\end{pmatrix}.$$
We find
$$\vert\vert \psi\vert\vert^2=\int {dz_1\wedge d\bar z_1\wedge
dz_2\wedge d\bar z_2 \over y^{3}}
\Biggl(\alpha^2\vert \psi_1+\psi_2\vert^2+
\alpha^{-2}\vert \beta(\psi_1+\psi_2)+\psi_2\vert^2\Biggr).$$
Therefore, $H^0(\mathbb{P}^2,\tilde V_1)=\mathbb{C}$ since there is only one section with finite norm
\be \label{triplet}\psi_{hol}=\begin{pmatrix}1\cr -1\end{pmatrix}.\ee
We further find $H^2(\mathbb{P}^2,\tilde V_1)=0$ 
since general harmonic (0,2) forms valued in $\tilde V_1$
are written as\footnote{We use that $*d\bar z_1\wedge d\bar z_2=d\bar z_1\wedge d\bar z_2$ in solving $\overline{D}^{\dg}\psi_{unit}=0.$}
$$\psi_{unit}=d\bar z_1\wedge d\bar z_2 \Bigl( \mathcal{G}^{-1}\Bigr)^{\dg}\begin{pmatrix}\psi_1\cr \psi_2\end{pmatrix}$$
and the norm
$$\vert\vert \psi\vert\vert^2=\int dz_1\wedge d\bar z_1\wedge
dz_2\wedge d\bar z_2 
\Biggl( \vert \alpha^{-1}(1+\overline{\beta})\psi_1-\alpha \psi_2  \vert^2+
\vert -\alpha^{-1}\overline{\beta} \psi_1+\alpha \psi_2\vert^2\Biggr)$$
diverges for any holomorphic $\psi_1,\psi_2.$
Then from the holomorphic Euler characteristic $\chi(\mathbb{P}^2,\tilde V_1)=1$
we find
$H^1(\mathbb{P}^2,\tilde V_1)=0.$
We will use this information in \cite{S} to identify $\tilde V_1$ as
a certain well-known holomorphic vector bundle on $\mathbb{P}^2.$

If one would like to compute the OPE (\ref{goal}),  the bundles $\mathcal{V}$ on $\mathcal{M}$ and  $\mathcal{V}_{bulk}$ on $\cM_{bulk},$ which appear
in (\ref{bub_intro}), are the pull-back of the vector bundle $\mathbf{V}_{a,b}$
over the base $\mathbb{P}^2_{\vec U}$ to $\cM$ and $\cM_{bulk}$ correspondingly. Indeed, recall that the total moduli space $\cM$ is $\mathbb{P}^2$ fibration over $\mathbb{P}^2_{\vec U}$
where the base (the fiber) is the space of Hecke modifications corresponding to
the first (the second) WH operator in the OPE. The bundles 
$\mathcal{V}$ and $\mathcal{V}_{bulk}$ are the pull-back from the base since only the first Wilson-'t Hooft operator
in the left side of (\ref{goal}) carries non-zero electric weight.

Since $\cM$ is compact we can take a connection on $\cV$ to be
the pull-back of the connection on the base $\mathbb{P}_{\vec U}^2.$
We will clarify how to define a connection $\cV_{bulk}$ on the non-compact $\cM_{bulk}$ in \cite{S} where we will present the
computation of the OPE (\ref{goal}).

\section{Conclusion}
In this note we determined the basic ingredients required to
compute the OPE (\ref{goal}) of Wilson-'t Hooft loop operators
in $\cN=4$ SYM theory with gauge group $G=PSU(3).$
This work is an extension of our approach \cite{KS} which
uses the holomorphic-topological twist  \cite{htft} of
the $\cN=4$ SYM theory and the connection between BPS configurations in $\cN=4$ SYM theory 
in the presence of 't Hooft operators and solutions of 3d Bogomolny equations 
with magnetic sources \cite{KW},\cite{Witten2009}.

In Section \ref{total} we found the compact moduli space  $\mathcal{M}$
of BPS configurations in the theory on $R\times I \times \mathcal{C}$ with two
't Hooft operators $W_{\mu,0}$ inserted at points in $I\times \mathcal{C}.$
The $PSU(3)$ invariant K\"ahler form on $\cM$ is written in (\ref{metric_tot})
with functions $f_i(s)$ given in (\ref{choice-tot}). We further determined
the non-compact space $\cM_{bulk}$ by removing from $\cM$ the
vicinity of the blown-up region corresponding to the bubbled contribution.
The $PSU(3)$ invariant K\"ahler form on $\cM_{bulk}$ is written in (\ref{metric})
with functions $g_i(s)$ given in (\ref{choice-bulk}). 

We computed $L^2$ Dolbeault cohomology
of $\cM$ and $\cM_{bulk}$ in Appendix A and Appendix B respectively.
This allowed us to verify our results about geometry of these moduli
spaces by making consistency check. Namely, we verified the OPE
of 't Hooft operators (\ref{answer}), predicted by S-duality, by
making explicit the action of principle $SU(2)$ subgroup of the dual group
 $\LG=SU(3)$ on the cohomology. This is in agreement with general
 facts about  moduli spaces of BPS configurations in the presence
 of 't Hooft operators \cite{Witten2009}.  

We further determined the vector bundles $\mathcal{V}$ and
$\mathcal{V}_{bulk}$ in Section \ref{bundle}. These bundles
take into account electric degrees of freedom present in dyonic operators
 in the OPE (\ref{goal}). We will compute the right side of (\ref{goal})
 for $\cN=4$ SYM and $\cN=2$ SYM with $N_f=0$ in the future \cite{S}
 and hope to compare with the forthcoming results from the
 alternative method \cite{GMN2} based on the connection with 2d CFT.

\section*{Acknowledgments}

I would like to thank F. Cachazo, J. Gomis, A. Kapustin, S. Katz,
R. Moraru,  T. Okuda, M. Rocek  for
discussions. I am especially grateful to A. Kapustin and S. Katz for
valuable advices. This research was supported in part by DARPA under Grant No. 
HR0011-09-1-0015 and by the National Science Foundation under Grant No. 
PHY05-51164. This material is based upon work supported in part by the National Science Foundation under Grant No. 1066293 and the hospitality of the Aspen Center for Physics.

\startappendix

\section{Appendix: $L^2$ Dolbeault Cohomology of $\cM$}
Here we compute cohomology groups $H^p(\cM,\Omega^p)$
for $p=0,1,2,3,4.$ It is clear that $h^p(\cM,\Omega^q)=0$ for $p\ne q$ as
follows from the non-trivial transformation of the basic
differentials (\ref{proper}) under the two $C^*$ actions (\ref{action}).
We could have computed cohomology groups of $\cM$ simply using
that $\cM$ is $\mathbb{P}^2$ fibration over $\mathbb{P}^2$. Instead, we chose to
compute  $L^2$ Dolbeault cohomology  using K\"ahler form on $\cM$ to verify that we have correctly identified the
geometry of $\cM.$  Moreover, this allows us to directly compare with
the $L^2$ Dolbeault cohomology of $\cM_{bulk},$ which we compute
in Appendix B, and identify the bubbled contribution (see Section \ref{check}).

\subsection{Harmonic (0,0) and (4,4) forms}
To compute the volume of $\mathcal{M},$ we introduce polar coordinates:
$$\lambda=\vert \lambda \vert e^{i\varphi},\quad z_1=r_1e^{\phi_1},\quad z_2=r_2e^{\phi_2},\quad v=r_ve^{\phi_v}$$
and write
$$x=a+b cos \Phi,\quad  a=1+t_2+t_v(1+t_1),\quad b=2r_1r_2r_v ,\quad
\Phi=\phi_2-\phi_1-\phi_v,\quad y=1+t_1+t_2$$
$$t_v=r_v^2,\quad t_1=r_1^2,\quad t_2=r_2^2.$$

Using explicit expressions (\ref{differ_ii}) for differentials we compute
\be \label{vol}\mathcal{E}_1\wedge\mathcal{E}_2 \wedge \mathcal{E}_3 \wedge \mathcal{E}_4\wedge\overline{\mathcal{E}}_1\wedge
\overline{\mathcal{E}}_2 \wedge \overline{\mathcal{E}}_3 \wedge
\overline{\mathcal{E}}_4
=\half s ds \wedge d\varphi \,\wedge {dv\wedge d\bar v\over x^2}\,\,
 \wedge {dz^1 \wedge d\bar z_1\wedge dz^2 \wedge d\bar z_2\over  y^2} \ee
 Then the volume of $\mathcal{M}$ is given by
$${\rm vol}_{\mathcal{M}}={(2\pi)^4\over 2}\int_{0}^{\infty} s\,ds\, f_1\,f_2\,f_3\,f_4$$
where we used
the following  integrals over $\Phi$ and $t_v$:
$$\int_{0}^{2\pi}{d\Phi\over \bigl(a+b cos \Phi\bigr)^2}={2\pi a \over (a^2-b^2)^{3/2}},\quad a>b$$
$$\int_{0}^{\infty}{t_v dt_v \over \bigl(\beta t_v^2+2\gamma t_v+\delta\bigr)^{3/2}}=\half {1\over y(1+t_1)},\quad
\int_{0}^{\infty}{dt_v \over \bigl(\beta t_v^2+2\gamma t_v+\delta\bigr)^{3/2}}=\half {1\over y(1+t_2)}$$
where
$$\beta=(1+t_1)^2,\quad \gamma=1+t_1+t_2-t_1t_2,\quad \delta=(1+t_2)^2.$$
Using asymptotics (\ref{zero}) and (\ref{inf}) of $f_i(s),$ we find that volume form on $\mathcal{M}$ is convergent i.e. both $(0,0)$ form $(4,4)$ forms are square
integrable.

\subsection{Harmonic (1,1) and (3,3) forms}
General $PSU(3)$ invariant (1,1) form is written as:
$$\omega=a_1 \mathcal{E}_1 \wedge \overline{\mathcal{E}}_1+
a_2 \mathcal{E}_2 \wedge \overline{\mathcal{E}}_2+
a_3 \mathcal{E}_3 \wedge
\overline{\mathcal{E}}_3+a_4 \mathcal{E}_4 \wedge \overline{\mathcal{E}}_4 .$$

Taking the Hodge star operation we find:
$$*{\omega}=c_1 \, \mathcal{E}_2 \wedge \mathcal{E}_3 \wedge
 \mathcal{E}_4 \wedge
\overline{\mathcal{E}}_2 \wedge \overline{\mathcal{E}}_3 \wedge
\overline{\mathcal{E}}_4 +
c_2\,  \mathcal{E}_1 \wedge \mathcal{E}_3\wedge \mathcal{E}_4
\wedge \overline{ \mathcal{E}}_1 \wedge
\overline{\mathcal{E}}_3 \wedge \overline{\mathcal{E}}_4
+c_3 \, \mathcal{E}_1 \wedge \mathcal{E}_2\wedge \mathcal{E}_4
 \wedge \overline{\mathcal{E}}_1 \wedge
\overline{\mathcal{E}}_2 \wedge \overline{\mathcal{E}}_4 +$$
$$c_4 \,  \mathcal{E}_1 \wedge \mathcal{E}_2\wedge \mathcal{E}_3 \wedge
\overline{ \mathcal{E}}_1\wedge
\overline{\mathcal{E}}_2 \wedge \overline{\mathcal{E}}_3$$

where
$$c_i=-a_i B_i \quad i=1,\ldots,4$$
$$B_1={f_2f_3f_4\over f_1}, \quad
B_2={f_1f_3f_4\over f_2},\quad B_3={f_1f_2f_4\over f_3},
\quad B_4={f_1f_2f_3\over f_4}.$$
To be square-integrable, $\omega$ must satisfy
\be \label{norm}\int \omega \wedge *\overline{\omega}={(2\pi)^4\over 2} \int_{0}^{\infty}ds\, s\Bigl(a_1^2B_1+
a_2^2B_2+a_3^2B_3
+a_4^2B_4\Bigr)< \infty\ee

We find that $\dbar \omega=0$ implies:
$$a_3'=-sa_1,\quad a_4=\#-2a_3,\quad a_2=a_3+a_4$$
where $\#$ is an integration constant.
Next, $\partial * {\omega}=0$ gives:
$$-sc_2+sc_3-2sc_4+c_1'=0.$$
For the choice of asymptotics (\ref{zero}) at $s \mapsto 0$
 we find
that all three solutions behave like constants
at $s\mapsto 0,$ which ensures that each of them gives finite contribution to the norm of the solution
from integrating around $s=0$.

At $s\mapsto \infty$ we use (\ref{inf}) to find general solution
$$a_3\sim \delta_1 s^{-1}+\delta_2 s +\delta_3 s^{-2}$$
$$a_4=-\delta_1 {C_1^2+3C_1C_2+2C_2^2\over C_2^2}
-\delta_3 {(C_1+2C_2)^2\over C_2^2}-2a_3$$
One has to set $\delta_2=0$ to ensure convergence
of  the integral in the definition of the norm. 
We conclude that the vector space of harmonic square-integrable (1,1) forms is
two dimensional. As a basis in this space, we can take the K\"ahler form
$\omega_{(1,1)}^{(1)}=J_{tot}$ and the form $\omega_{(1,1)}^{(2)}$
orthogonal to $J_{tot}$ i.e.
such that
\be \label{orth}J_{tot}\wedge * \omega_{(1,1)}^{(2)}=0.\ee
Note that $\omega_{(1,1)}^{(1)}$ corresponds to $\delta_1=-\delta_3=C_2$
while  $\omega_{(1,1)}^{(2)}$ to $\delta_1=C_1+2C_2,\,\delta_3=-C_1.$

By Serre duality the space of  harmonic square-integrable (3,3) forms is also
two dimensional. As a basis, we may take
$$\omega_{(3,3)}^{(1)}=J_{tot}^3,\quad \omega_{(3,3)}^{(2)}=J_{tot}^2\wedge \omega_{(1,1)}^{(2)}$$

\subsection{Harmonic (2,2) forms}
General $PSU(3)$ invariant (2,2) form is written as:
$$\omega=h_1 \mathcal{E}_1 \wedge \mathcal{E}_3
\wedge \overline{\mathcal{E}}_1\wedge \overline{\mathcal{E}}_3
+h_2 \mathcal{E}_1 \wedge \mathcal{E}_4
\wedge \overline{\mathcal{E}}_1\wedge \overline{\mathcal{E}}_4
+h_3  \mathcal{E}_1 \wedge \mathcal{E}_2
\wedge \overline{\mathcal{E}}_1\wedge \overline{\mathcal{E}}_2+
h_4 \mathcal{E}_1 \wedge \mathcal{E}_2
\wedge \overline{\mathcal{E}}_3\wedge \overline{\mathcal{E}}_4+$$
$$h_5 \mathcal{E}_3 \wedge \mathcal{E}_4
\wedge \overline{\mathcal{E}}_1\wedge \overline{\mathcal{E}}_2+
h_6 \mathcal{E}_3 \wedge \mathcal{E}_4
\wedge \overline{\mathcal{E}}_3\wedge \overline{\mathcal{E}}_4+
h_7 \mathcal{E}_2 \wedge \mathcal{E}_3
\wedge \overline{\mathcal{E}}_2\wedge \overline{\mathcal{E}}_3+
h_8 \mathcal{E}_2 \wedge \mathcal{E}_4
\wedge \overline{\mathcal{E}}_2\wedge \overline{\mathcal{E}}_4
$$
From $\overline{\partial} \omega=0$ we find:
$$s h_4'+h_4+s(h_3-h_2- h_1)=0,\quad
h_6'-2sh_1+sh_2+h_5=0,$$
\be \label{two-two}h_7'-sh_1+sh_3-h_5=0,\quad
h_8'-sh_2- h_5-2 s h_3=0.
\ee

\subsubsection{Self-dual  forms}
Let us first look  for self-dual (2,2) forms solving (\ref{two-two})
\be\label{self-tot}h_4=h_5,\quad h_1=h_8 A_{24},\quad
 h_2=h_7 A_{23}  \quad
 h_3=h_6 A_{34}\ee
 where
 $$A_{24}={f_1f_3\over f_2f_4},\quad A_{23}={f_1f_4\over f_2f_3},
\quad A_{34}= {f_1f_2\over f_3f_4}.$$
To be square-integrable, $\omega$ must satisfy
\be \label{norm2}
\int \omega \wedge *\overline{\omega}={(2\pi)^4\over 2}\int ds\, s \Bigl(h_4^2+h_6^2A_{34}+
h_7^2A_{23}+
h_8^2A_{24}\Bigr)< \infty\ee
There is one obvious square-integrable solution
$$\omega_{(2,2)}=J_{tot}\wedge J_{tot}.$$

Let us look for other solutions among primitive self-dual forms
i.e. we impose $J_{tot}\wedge \omega_{(2,2)}=0$ which, using self-duality
(\ref{self-tot}), amounts to
\be \label{prim-tot}h_7=-h_6-f_3H,\quad h_8=-h_6+f_4H.\ee
Then, equations (\ref{two-two}) reduce to three ODEs
for three functions $h_4,h_6,H$:
$$
f_4H'+2sf_1H-2s\bigl(A_{34}-A_{24}\bigr)h_6-2sf_4 A_{24}H=0,
$$
\be \label{prim}h'_6+sh_6\bigl(2A_{24}- A_{23}\bigr)+h_4-sH\bigl(f_3A_{23}+2f_4 A_{24}\bigr)=0,\ee
$$sh'_4+h_4+sh_6\Bigl(A_{34}+ A_{23}+A_{24}\Bigr)+
sH\Bigl(f_3A_{23}-f_4 A_{24}\Bigr)=0.$$

Using (\ref{inf}) we find general solution of (\ref{prim-tot})  at $s\mapsto \infty$:
 $$h_4=-\delta_1 s+\frac{(2\delta_2 + 3\delta_3)}{s^2}
 \quad
h_6=\delta_1 s^2 +\frac{(\delta_2+\delta_3)}{s}\quad
H=\frac{\delta_1}{2}s^2-\frac{\delta_2}{s}+2\delta_3
$$
where $\delta_i$ are constants.
We see that among primitive forms, there are two well-behaved at $s\mapsto \infty $ solutions
obtained by choosing $\delta_1=0.$

Using (\ref{zero}) we find general solution of (\ref{prim-tot}) at $s\mapsto 0$:
 $$h_4= \kappa_1 s^{-1} -\kappa_2 {C_2(C_1-C_2)\over C_1+C_2}
 -2\kappa_3 {(C_1^2+C_1C_2+C_2^2)\over C_1(C_1+C_2)}$$
$$h_6=-\kappa_1 Log(s)+ \kappa_3+a(\kappa_2,\kappa_3)s, $$
$$H=-2{(C_1+2C_2)\over C_1(C_1+C_2)}\kappa_1 s Log(s)+
\kappa_2 + b(\kappa_2,\kappa_3)s$$
where $a,b$ are linear combinations of constants $\kappa_2$
and $\kappa_3.$ Recall that $C_1,C_2$ are K\"ahler moduli
which appear in $J_{tot}$ (\ref{choice-tot}).
Setting $\kappa_1=0$  leaves 2 solutions well-behaved at
$s\mapsto 0.$

We checked using Mathematica that each
of the two good solutions of (\ref{prim-tot}) at $s \mapsto 0$ (parametrized
by $\kappa_2,\kappa_3$)  interpolates
at $s\mapsto \infty$ into a bad solution with $\delta_1\ne 0.$
There is a linear combination of the two good solutions
at  $s \mapsto 0$ which interpolates into a good solution at $s\mapsto \infty.$ 
Therefore the space of primitive self-dual harmonic (2,2) forms is one dimensional.
In total, we conclude that the space of  self-dual harmonic (2,2) forms on $\cM$ is two dimensional and we may choose as basis vectors 
$\omega_{(2,2)}^{(1)}=J_{tot}^2$ and $\omega_{(2,2)}^{p.s.d}$ such that
$$\omega^{p.s.d}_{(2,2)}=*\omega^{p.s.d}_{(2,2)}\quad \text{and} \quad \omega^{p.s.d}_{(2,2)}\wedge J_{tot}=0.$$

\subsubsection{Anti-self-dual forms}

Let us now look  for anti-self-dual (2,2) forms solving (\ref{two-two})
$$h_4=-h_5,\quad h_1=-h_8 A_{24},\quad
 h_2=-h_7 A_{23}  \quad
 h_3=-h_6 A_{34}.$$

There is an obvious solution:
\be \label{obv}\omega^{a.s.d}_{(2,2)}=J_{tot}\wedge \omega^{(2)}_{(1,1)}\ee
where the form $\omega^{(2)}_{(1,1)}$
appeared  in Section 3.2. This (1,1) form
is orthogonal to $J_{tot},$ see (\ref{orth}), which ensures the
anti-self-duality of $\omega^{a.s.d}_{(2,2)}.$

Let us prove that the space of harmonic
square-integrable anti-self-dual (2,2) forms is one dimensional.
Using (\ref{inf}) we find  general solution at $s\mapsto \infty$
 $$h_4=\delta_1+\delta_3 s^{-3},\quad
 h_6=(\delta_1+\delta_2)s+(\delta_4-\delta_3)s^{-2}$$
$$h_7=\delta_2 s+\delta_3 s^{-2},\quad 
h_8=-(3\delta_1+3\delta_2)s+\bigl(16+{1\over s^2}\bigr)\delta_4$$
where $\delta_i$ are constants. Setting $\delta_1=\delta_2=0$ leaves 2 solutions well-behaved at
$s\mapsto \infty.$

 Meanwhile, with $s\mapsto 0$  asymptotics (\ref{zero}) we find
  three well-behaved solutions
$$h_4=\gamma h_6^{(0)}-\beta h_7^{(0)}-\alpha h_8^{(0)}+O(s)$$
$$h_6=h_6^{(0)}+O(s),\quad h_7=h_7^{(0)}+O(s),\quad h_8=h_8^{(0)}+O(s).$$
where $h_i^{(0)}$ for $i=6,7,8$ are independent constants.
The fourth solution is not well-behaved:
$$h_4={\delta \over s}+O\bigl(ln(s)\bigr),\quad h_6,h_7,h_8\sim O\bigl(ln(s)\bigr).$$

 We use Mathematica to show that out of 3 solutions
 well-behaved at $s\mapsto 0$ we can construct only one
 linear combination which also behaves well at $s\mapsto \infty.$
 More concretely,   each of the three solutions, parametrized by $h_k^{(0)}$
 with $k=6,7,8,$  interpolates to a solution with non-zero
 $\delta_1$ and $\delta_2$  at large $s$. We can construct only one
 linear combination of the three good solutions at $s\mapsto 0$
 which interpolates to a solution with $\delta_1=\delta_2=0$ at $
 s\mapsto \infty.$
We conclude that the space of square-integrable anti-self-dual harmonic (2,2) forms on $\cM$ is one dimensional with a basis vector (\ref{obv}).

\section{Appendix: $L^2$ Dolbeault Cohomology of $\cM_{bulk}$}
Here we compute cohomology groups $H^p(\cM_{bulk},\Omega^p)$
for $p=0,1,2,3,4.$ It is clear that $h^p(\cM_{bulk},\Omega^q)=0$ for $p\ne q$ as
follows from the non-trivial transformation of the basic
differentials (\ref{proper}) under the two $C^*$ actions (\ref{action}).

\subsection{Harmonic (0,0) and (4,4) forms}
The volume of $\mathcal{M}_{bulk}$ is computed in the same
way as the volume of $\mathcal{M}$ with substitution $f_i(s)\mapsto g_i(s).$ We find
$${\rm vol}_{\mathcal{M}_{bulk}}={(2\pi)^4\over 2}\int_{0}^{\infty} s\,ds\, g_1\,g_2\,g_3\,g_4.$$

Using asymptotics (\ref{inf_bulk}) and (\ref{zero_bulk}) of $g_i(s)$ we find that volume form on $\mathcal{M}_{bulk}$ is convergent i.e. both $(0,0)$ form $(4,4)$ forms are in  $L^2.$

\subsection{Harmonic (1,1) and (3,3) forms}
General $PSU(3)$ invariant (1,1) form is written as:
$$\omega_{(1,1)}=a_1 \mathcal{E}_1 \wedge \overline{\mathcal{E}}_1+
a_2 \mathcal{E}_2 \wedge \overline{\mathcal{E}}_2+
a_3 \mathcal{E}_3 \wedge
\overline{\mathcal{E}}_3+a_4 \mathcal{E}_4 \wedge \overline{\mathcal{E}}_4 .$$
From $\delbar \omega_{(1,1)}=0$ and $\delbar^{\dagger}\omega_{(1,1)}=0$ we find
$$a_3'=-sa_1,\quad a_4=\# -2a_3,\quad a_2=a_3+a_4$$
$$-{\hat B_1\over s} a_3''-\left({\hat B_1\over s}\right)'a_3'+
s(\hat B_2+\hat B_3+4\hat B_4)a_3=s \#(\hat B_2+2\hat B_4)$$
where $\hat B_i={\prod_{j\ne i}g_j(s) \over g_i(s)}$ and $\#$ is a constant.
To be square-integrable, $\omega_{(1,1)}$ must satisfy
\be \label{norm_bulk}\int \omega_{(1,1)} \wedge *\overline{\omega}_{(1,1)}={(2\pi)^4\over 2} \int_{0}^{\infty}ds\, s\Bigl(a_1^2\hat B_1+
a_2^2\hat B_2+a_3^2\hat B_3
+a_4^2\hat B_4\Bigr)< \infty\ee
There is an obvious square-integrable solution - the K\"ahler form on $\cM_{bulk}$
$$\omega_{(1,1)}=J_{bulk}$$
but let us look for other solutions.

For the choice of asymptotics (\ref{zero_bulk}) at $s \mapsto 0$
 we find that
 general solution  has the form:
$$a_3=\kappa_1 s^{1\over 2}+\kappa_2s^{-{1\over 2}}+
\kappa_3 $$
with integration constants $\kappa_1,\kappa_2,\kappa_3.$
There are two well-behaved solutions at $s \mapsto 0$
obtained by setting $\kappa_2=0.$ The K\"ahler form $J_{bulk}$
corresponds to further taking $\kappa_3=-\kappa_1=\hat A.$

Meanwhile, using (\ref{inf_bulk}), general solution at $s \mapsto \infty$
has the form:
$$a_3=\gamma_1 s^{-1}+\gamma_2 s+\gamma_3s^{-2}.$$
There are two well-behaved solutions at $s \mapsto \infty$
obtained by setting $\gamma_2=0$.

Using Mathematica we checked that if $\kappa_3\ne -\kappa_1,$
then a good solution at $s\mapsto 0$ interpolates into a bad solution
with $\gamma_2 \ne 0$ at $s \mapsto \infty.$ 

We conclude that the vector space of harmonic square-integrable (1,1) forms
on $\cM_{bulk}$ is one dimensional with a basis vector
 $J_{bulk}$ .
By Serre duality we also get that the space of harmonic square-integrable (3,3) forms
is one dimensional with a basis vector
$J_{bulk}^3.$

\subsection{Harmonic (2,2) forms}

General $PSU(3)$ invariant (2,2) form is written as:
$$\omega_{(2,2)}=h_1 \mathcal{E}_1 \wedge \mathcal{E}_3
\wedge \overline{\mathcal{E}}_1\wedge \overline{\mathcal{E}}_3
+h_2 \mathcal{E}_1 \wedge \mathcal{E}_4
\wedge \overline{\mathcal{E}}_1\wedge \overline{\mathcal{E}}_4
+h_3  \mathcal{E}_1 \wedge \mathcal{E}_2
\wedge \overline{\mathcal{E}}_1\wedge \overline{\mathcal{E}}_2+
h_4 \mathcal{E}_1 \wedge \mathcal{E}_2
\wedge \overline{\mathcal{E}}_3\wedge \overline{\mathcal{E}}_4+$$
$$h_5 \mathcal{E}_3 \wedge \mathcal{E}_4
\wedge \overline{\mathcal{E}}_1\wedge \overline{\mathcal{E}}_2+
h_6 \mathcal{E}_3 \wedge \mathcal{E}_4
\wedge \overline{\mathcal{E}}_3\wedge \overline{\mathcal{E}}_4+
h_7 \mathcal{E}_2 \wedge \mathcal{E}_3
\wedge \overline{\mathcal{E}}_2\wedge \overline{\mathcal{E}}_3+
h_8 \mathcal{E}_2 \wedge \mathcal{E}_4
\wedge \overline{\mathcal{E}}_2\wedge \overline{\mathcal{E}}_4
$$
From $\overline{\partial} \omega=0$ we find:
$$s h_4'+h_4+s(h_3-h_2- h_1)=0,\quad
h_6'-2sh_1+sh_2+h_5=0,$$
\be \label{two-two_bulk}h_7'-sh_1+sh_3-h_5=0,\quad
h_8'-sh_2- h_5-2 s h_3=0.
\ee

\subsubsection{Self-dual  forms}
Let us first look  for self-dual (2,2) forms solving (\ref{two-two_bulk})
\be \label{self}h_4=h_5,\quad h_1=h_8 \hat A_{24},\quad
 h_2=h_7 \hat A_{23}  \quad
 h_3=h_6 \hat A_{34}\ee
 where
 $$\hat A_{24}={g_1g_3\over g_2g_4},\quad \hat A_{23}={g_1g_4\over g_2g_3},
\quad \hat A_{34}= {g_1g_2\over g_3g_4}.$$
To be square-integrable, $\omega$ must satisfy
\be \label{norm2_bulk}
\int \omega_{(2,2)} \wedge *\overline{\omega}_{(2,2)}={(2\pi)^4\over 2} \int ds\, s \Bigl(h_4^2+h_6^2\hat A_{34}+
h_7^2\hat A_{23}
+h_8^2\hat A_{24}\Bigr)< \infty\ee
There is one obvious square-integrable solution
$$\omega_{(2,2)}=J_{bulk}\wedge J_{bulk}.$$

Let us look for other solutions among primitive self-dual forms
i.e. we impose $J_{bulk}\wedge \omega=0$ which, using self-duality
(\ref{self}), amounts to
$$h_7=-h_6-g_3H,\quad h_8=-h_6+g_4H.$$
Then, equations (\ref{two-two_bulk}) reduce to three ODEs
for three functions $h_4,h_6,H$:
$$
g_4H'+2sg_1H-2s\bigl(\hat A_{34}-\hat A_{24}\bigr)h_6-2sg_4 \hat A_{24}H=0,
$$
\be \label{prim-bulk}h'_6+sh_6\bigl(2\hat A_{24}-\hat A_{23}\bigr)+h_4-sH\bigl(g_3\hat A_{23}+2g_4\hat A_{24}\bigr)=0,\ee
$$sh'_4+h_4+sh_6\Bigl(\hat A_{34}+\hat A_{23}+\hat A_{24}\Bigr)+
sH\Bigl(g_3\hat A_{23}-g_4\hat A_{24}\Bigr)=0.$$

Using (\ref{inf_bulk}) we find general solution of (\ref{prim-bulk})  at $s\mapsto \infty$:
 $$h_4=-\delta_1 s+\frac{(2\delta_2 + 3\delta_3)}{s^2}
 \quad
h_6=\delta_1 s^2 +\frac{(\delta_2+\delta_3)}{s}\quad
H=\frac{\delta_1}{2}s^2-\frac{\delta_2}{s}+2\delta_3
$$
where $\delta_i$ are constants.
We see that among primitive forms, there are two well-behaved at $s\mapsto \infty $ solutions
obtained by choosing $\delta_1=0.$

Using (\ref{zero_bulk}) we find general solution of (\ref{prim-bulk}) at $s\mapsto 0$:
 $$h_4=-\half \tC_1 s^{-2} -
 \half \tC_2 s^{-\half}-3\tC_3,\,\,
h_6=-\tC_1 s^{-1}+
\half \tC_2 s^{1\over 2}+\tC_3\bigl(s^{1\over 2}+2s\bigr),$$
$$H=\tC_1s^{-1}+\tC_2s^{\half}+\tC_3.$$
Setting $\tC_1=0$  leaves 2 solutions well-behaved at
$s\mapsto 0.$

We checked using Mathematica that each
of the two good solutions of (\ref{prim-bulk}) at $s \mapsto 0$ (parametrized
by $\tC_2,\tC_3$)  interpolates
at $s\mapsto \infty$ into a bad solution with $\delta_1\ne 0.$
There is a linear combination of the two good solutions
at  $s \mapsto 0$ which interpolates into a good solution at $s\mapsto \infty.$ 
Therefore the space of primitive self-dual harmonic (2,2) forms is one dimensional.
In total, we conclude that the space of  self-dual harmonic (2,2) forms on $\cM_{bulk}$ is two dimensional and we may choose as basis vectors 
$\omega_{(2,2)}^{(1)}=J_{bulk}^2$ and $\omega_{(2,2)}^{p.s.d}$ such that
$$\omega^{p.s.d}_{(2,2)}=*\omega^{p.s.d}_{(2,2)}\quad \text{and} \quad \omega^{p.s.d}_{(2,2)}\wedge J_{bulk}=0.$$

\subsubsection{Anti-self-dual forms}
Let us now look  for anti-self-dual (2,2) forms solving (\ref{two-two_bulk})
$$h_4=-h_5,\quad h_1=-h_8 \hat A_{24},\quad
 h_2=-h_7 \hat{A}_{23}  \quad
 h_3=-h_6 \hat{A}_{34}$$
Using (\ref{inf_bulk}) we find  general solution at $s\mapsto \infty$
 $$h_4=\delta_1+\delta_3 s^{-3},\quad
 h_6=(\delta_1+\delta_2)s+(\delta_4-\delta_3)s^{-2}$$
$$h_7=\delta_2 s+\delta_3 s^{-2},\quad 
h_8=-(3\delta_1+3\delta_2)s+\bigl(16+{1\over s^2}\bigr)\delta_4$$
where $\delta_i$ are constants. Setting $\delta_1=\delta_2=0$ leaves 2 solutions well-behaved at
$s\mapsto \infty.$

Using (\ref{zero_bulk}) we find general solutions at $s \mapsto 0$:
 $$h_4=\tC_1+\tC_3 s^{-3/2},\quad
 h_6=2(\tC_1+\tC_2)s+\tC_4 s^{-1/2}$$
$$h_7=\tC_2(1+s)+3{\tC_3}s^{-1/2},\quad 
h_8=-2(\tC_1+\tC_2)s+(2\tC_3+\tC_4)s^{-1/2}$$
where  $\tC_i$  are constants.
 Setting $\tC_3=\tC_4=0$ leaves 2 solutions well-behaved at
$s\mapsto 0.$

We checked using Mathematica that there are no square integrable harmonic anti-selfdual (2,2) forms on $\cM_{bulk}.$ Namely,
each of the two good solutions at $s \mapsto \infty$
interpolates to a solution at $s\mapsto 0$ with both $\tC_3\ne 0$ and $\tC_4\ne 0.$
It is not possible to eliminate these divergent
pieces at $s\mapsto 0$   by any linear combination of
the two good solutions at $s\mapsto \infty.$

\end{document}